\documentclass[fleqn,usenatbib]{mnras}

\usepackage{newtxtext,newtxmath}

\usepackage[T1]{fontenc}

\DeclareRobustCommand{\VAN}[3]{#2}
\let\VANthebibliography\thebibliography
\def\thebibliography{\DeclareRobustCommand{\VAN}[3]{##3}\VANthebibliography}

\usepackage{graphicx}	
\usepackage{amsmath}	
\usepackage{orcidlink}

\usepackage{booktabs}

\title[The ECLE AT 2022fpx]{The extreme coronal line emitter AT 2022fpx: Varying optical polarization properties and late-time X-ray flare}

\author[K. I. I. Koljonen et al.]{Karri I. I. Koljonen$^{\orcidlink{0000-0002-9677-1533}}$,$^{1}$\thanks{E-mail: karri.koljonen@ntnu.no}
Ioannis Liodakis$^{\orcidlink{0000-0001-9200-4006}}$,$^{2,3,4}$
Elina Lindfors$^{\orcidlink{0000-0002-9155-6199}}$,$^{2,5}$
Kari Nilsson,$^{2}$
\newauthor
Thomas M. Reynolds,$^{6,5}$ 
Panos Charalampopoulos$^{\orcidlink{0000-0002-0326-6715}}$,$^{5}$
Konstantinos Kouroumpatzakis$^{\orcidlink{0000-0002-1444-2016}}$,$^{7}$
\newauthor
Callum McCall,$^{8}$
Helen E. Jermak,$^{\orcidlink{0000-0002-1197-8501}}$,$^{8}$
Iain A. Steele$^{8}$
and Juan Carbajo-Hijarrubia$^{\orcidlink{0000-0002-2010-7513}}$,$^{9,10,11}$
\\
$^{1}$Institutt for Fysikk, Norwegian University of Science and Technology, H{\o}gskloreringen 5, Trondheim, 7491, Norway\\
$^{2}$Finnish Centre for Astronomy with ESO (FINCA), University of Turku, Finland \\
$^{3}$NASA Marshall Space Flight Center, Huntsville, AL 35812, USA\\
$^{4}$Institute of Astrophysics, Foundation for Research and Technology - Hellas, Heraklion, GR7110, Greece\\
$^{5}$Department of Physics and Astronomy, University of Turku, FI-20014 Turku, Finland\\
$^{6}$Cosmic Dawn Center (DAWN), Niels Bohr Institute, University of Copenhagen, 2200, Denmark\\
$^{7}$Astronomical Institute of the Czech Academy of Sciences,
Bocni II 1401, CZ-141 00, Praha 4 – Sporilov, Czech Republic\\
$^{8}$Astrophysics Research Institute, Liverpool John Moores University, Liverpool Science Park IC2, 146 Brownlow Hill, UK\\
$^{9}$Institut d'Estudis Espacials de Catalunya (IEEC), Gran Capit\`a, 2-4, 08034 Barcelona, Spain\\
$^{10}$Institut de Ci\`encies del Cosmos (ICCUB), Universitat de Barcelona (UB), Martí i Franquès, 1, 08028 Barcelona, Spain\\
$^{11}$Departament de Física Quàntica i Astrofísica (FQA), Universitat de Barcelona (UB),  Martí i Franquès, 1, 08028 Barcelona, Spain\\
}

\date{Accepted XXX. Received YYY; in original form ZZZ}

\pubyear{2015}

\begin{document}
\label{firstpage}
\pagerange{\pageref{firstpage}--\pageref{lastpage}}
\maketitle

\begin{abstract}
Supermassive black holes disrupt passing stars, producing outbursts called tidal disruption events (TDEs). TDEs have recently gained attention due to their unique dynamics and emission processes, which are still not fully understood. Especially, the so-called optical TDEs, are of interest as they often exhibit delayed or obscured X-ray emission from the accretion disk, making the origin of the prompt emission unclear. In this paper, we present multiband optical polarization observations and optical spectrometry of a recent TDE candidate AT 2022fpx, alongside monitoring observations in optical, ultraviolet and X-rays. The optical spectra of AT 2022fpx show Bowen fluorescence as well as highly-ionized iron emission lines, which are characteristic of extreme coronal line emitters. Additionally, the source exhibits variable but low-polarized continuum emission at the outburst peak, with a clear rotation of the polarization angle. X-ray emission observed approximately 250 days after the outburst peak in the decay appear flare-like but is consistent with constant temperature black-body emission. The overall outburst decay is slower than for typical TDEs, and resembles more the ones seen from Bowen fluorescence flares. These observations suggest that AT 2022fpx could be a key source in linking different long-lived TDE scenarios. Its unique characteristics, such as extreme coronal line emission, variable polarization, and delayed X-ray flare, can be attributed to the outer shock scenario or a clumpy torus surrounding the supermassive black hole. Further studies, especially in the context of multi-wavelength observations, are crucial to fully understand the dynamics and emission mechanisms of these intriguing astrophysical events.
\end{abstract}

\begin{keywords}
black hole physics -- galaxies: nuclei -- polarization -- shock waves -- X-rays: galaxies 
\end{keywords}



\section{Introduction}

Tidal disruption event (TDE), where a supermassive black hole (SMBH) disrupts structurally a passing star causing a surge of matter to accrete into the black hole, powers a luminous outburst typically detected in the X-ray and/or optical wavelengths \citep{komossa99,donley02}. Recent optical monitoring programs, such as the Zwicky Transient Facility (ZTF), have significantly increased the number of known TDEs but, at the same time, challenged our understanding of their dynamics and emission processes \citep{gezari12,vanvelzen21}. Many of these `optical TDEs' show X-ray emission only years after the optical peak indicating delayed or obscured emission from the accretion disk and leaving the origin of the prompt emission unclear \citep{gezari21}.

The most striking features in the TDE optical spectra are broad helium and/or hydrogen emission lines \citep[e.g.,][]{arcavi14}. However, it has become clear that higher ionization lines also exist in the spectra of some TDEs. These include lines formed by Bowen fluorescence \citep{bowen28} in dense gaseous matter surrounding an ionizing extreme ultraviolet source \citep[e.g.,][]{netzer85,kastner90}, and have been identified in hydrogen and helium-rich TDEs \citep[e.g.,][]{blagorodnova19, onori19, vanvelzen21, charalampopoulos22, wevers22}, where they are thought to originate from a hot and compact UV/optical photosphere \citep{vanvelzen21}. However, Bowen fluorescence lines are also observed in the case of `rejuvenated SMBHs' \citep{trakhtenbrot19}. Subsequently, these events have been termed `Bowen Fluorescence Flares' \citep[BFFs;][]{makrygianni23}, encompassing several sources that exhibit long-lasting UV/optical emission persisting for several years, along with Bowen fluorescence lines (typically He {\sc II} Ly$\alpha$ $\lambda$3038 induced O {\sc III} line blend at 3754 \AA, 3757 \AA, and 3759 \AA, and O {\sc III} $\lambda$374.4 induced N {\sc III} lines at 4097 \AA, 4104 \AA, and 4640 \AA\/ are observed in the optical), and broad hydrogen and helium emission lines with velocities of a few thousand kilometers per second.

In rare cases, even higher ionization lines than those of the Bowen lines have been observed from TDEs in the form of highly-ionized, or `coronal', emission lines ([Fe {\sc XIV}] $\lambda$5303, [Fe {\sc VII}] $\lambda\lambda$5720, 6087, [Fe {\sc X}] $\lambda$6374, [Fe {\sc XI}] $\lambda$7892, and [Fe {\sc XIII}] $\lambda$10798; \citealt{onori22,short23}). The extreme ionization strength of these coronal lines suggests the presence of a strong extreme-UV or soft X-ray ionizing source within a gas-rich environment \citep[e.g., similar to narrow-line regions in active galactic nuclei; AGN;][]{cerqueira-campos21,gravity21}. Similar lines have been observed from the nuclear spectra of so-called extreme coronal line emitters \citep[ECLEs; ][]{komossa08,komossa09,wang11,wang12} -- galactic nuclei that seem otherwise quiescent and not active. Recent comparisons of ECLE and TDE properties suggest that ECLEs are likely echoes of unobserved past TDEs in these galaxies \citep{hinkle24,clark24,callow24}.  

Several scenarios have been proposed to explain the emission from optical TDEs. The prevailing ones include reprocessed X-ray emission by the obscuring, surrounding gas from the stellar debris \citep[e.g.,][]{roth16} or radiatively driven wind \citep[e.g.,][]{metzger16,parkinson22}, and tidal stream shock powered emission \citep{shiokawa15}. In the latter scenario, intersecting stellar streams near the pericenter of the stellar orbit form a nozzle shock that dissipates orbital energy \citep{bonnerot22}. However, this energy loss is likely not enough to produce rapid circularization of the flow and the subsequent formation of the accretion disk \citep{piran15,steinberg24}. Instead, the stellar flow collides with itself further away from the black hole \citep[$10^{15}-10^{16}$ cm; $\mathcal{O}(10^3$ R$_g$);][]{ryu20}, producing strong shocks that can produce emission in the optical band.

Polarization observations (both spectro- and photopolarimetry) of optical TDEs are scarce, with only a few significant detections available in the literature \citep{Wiersema2012,Wiersema2020,Lee2020,leloudas22,Patra2022,liodakis23}. These studies showed that in most cases TDEs have a polarization level of a few per cent that is overall constant in regions that are continuum-dominated and free of strong, depolarizing emission lines. Only in one case, a very high polarization degree was inferred \citep[up to $\sim$25\%;][]{liodakis23}. Thus, polarization observations hold the potential to distinguish between different accretion disk formation scenarios. In the reprocessing scenario, the polarization degree from an electron-scattering photosphere is low \citep[a maximum of $\sim$6\% in the most favorable conditions;][]{leloudas22}. Conversely, in the outer shock scenario, synchrotron emission arising from strong shocks can lead to a much higher polarization degree \citep[up to $\sim$25\%;][]{liodakis23}. Hybrid scenarios, based on collision-induced outflows, have also been proposed to explain the observed polarization behavior \citep{Charalampopoulos2023}. However, the diverse polarization signatures found even in the limited number of TDEs studied so far present a challenge for current models to comprehensively explain.

\subsection{AT 2022fpx} \label{sec:fpx}

The transient AT 2022fpx (ATLAS22kjn, ZTF22aadesap, Gaia22cwy) was discovered on March 31, 2022 (MJD 59669) by the Asteroid Terrestrial-impact Last Alert System (ATLAS) with a discovery magnitude of 18.54 in the ATLAS/orange filter \citep{tonry22}. It was subsequently classified as a TDE candidate at a redshift of $z = 0.073$ \citep{perezfournon22}\footnote{\url{https://www.wis-tns.org/object/2022fpx}}. The event is associated with the nuclear region of the galaxy SDSS J153103.70$+$532419.3. The peak r- and g-band magnitudes of 17.6 mag and 17.8 mag, respectively, were reached on July 24, 2022 (MJD 59784), approximately 115 days after the discovery. Prior to the transient event, the long-term magnitude level has remained constant at around 20.1 mag in the \textit{Gaia} g-band.

Based on optical monitoring of the event by several optical transient facilities, e.g., the Zwicky Transient Facility\footnote{\url{https://lasair-ztf.lsst.ac.uk/objects/ZTF22aadesap/}} (ZTF) and \textit{Gaia}\footnote{\url{http://gsaweb.ast.cam.ac.uk/alerts/alert/Gaia22cwy/}}, AT 2022fpx has remained relatively bright since the peak emission, showing a slow decay currently at 18.9 mag (r-band, 2024-05-17). The classification optical spectrum of AT 2022fpx presented by \citet{perezfournon22} in the Transient Name Server (TNS) exhibits strong but moderately broad hydrogen and ionized helium emission lines with maximum velocities of a few thousand km/s.

The light curve behavior and spectral properties of AT 2022fpx deviate from the typical optical TDE light curve evolution, which exhibit faster decays, bluer optical continua, and higher optical line velocities around $\sim$10$^4$ km/s \citep[e.g.,][]{vanvelzen21,charalampopoulos22}. However, it is noteworthy that the TNS spectrum was obtained relatively late in the event, approximately 85 days after the first detection of the transient. Consequently, it remains uncertain whether AT 2022fpx exhibited broader emission lines at early times. Also, at comparable times to the first spectrum for AT 2022fpx, there are TDEs that display similar line widths, such as ASASSN-14li \citep{brown17}. Overall, the observational characteristics of AT 2022fpx align more closely with the 'rejuvenated SMBH' scenarios \citep{trakhtenbrot19}, and thus, AGN contribution cannot be ruled out.

In this paper, we present an optical spectrum obtained from the Nordic Optical Telescope (NOT), optical polarization observations from both the NOT and the Liverpool Telescope, and UV/X-ray observations from the \textit{Neil Gehrels Swift Observatory} and \textit{XMM-Newton}, taken during the transient event AT 2022fpx, along with publicly available optical photometric data from ZTF. Section \ref{sec:obs} provides details on the observations and data reduction. In Section \ref{sec:results}, we present the optical spectrum of AT 2022fpx, showing highly-ionized emission lines, in addition to Bowen, hydrogen, and helium lines, which indicate the source being an extreme coronal line emitter. We also derive and discuss the host galaxy properties. Further, we present the optical polarization properties of AT 2022fpx, demonstrating clear variability, along with optical/UV monitoring throughout the event. Additionally, we discuss the X-ray monitoring results, which indicate a late-time flare and spectral properties consistent with black-body emission. Section \ref{sec:discuss} delves into the discussion of the variable polarization, in addition to drawing comparisons with recent well-sampled TDEs sharing similar properties to AT 2022fpx. Finally, Section \ref{sec:conclude} concludes the paper. We utilize a flat cosmology with $H_0 = 69.6$ km s$^{-1}$ Mpc$^{-1}$ for our calculations. 

\section{Observations} \label{sec:obs}

\subsection{Optical spectroscopy}

In order to confirm the classification of AT~2022fpx we obtained an optical spectrum with the Alhambra Faint Object Spectrograph and Camera (ALFOSC) at the NOT using the grism Gr4. The spectrum was reduced using the {\sc pypeit} package \citep{pypeit:joss_pub,pypeit:zenodo} and included overscan, bias and flat-field corrections, cosmic ray removal, spectral extaction, wavelength calibration using arc lamps; creation of a sensitivity function using a standard star observed on the same night; flux calibration with this sensitivity function; and telluric correction via fitting to a grid of atmospheric models.

\subsection{Optical polarization}

We observed AT 2022fpx using ALFOSC at the NOT in 13 epochs during the transient event from MJD 59772.5 to MJD 60173 with irregular cadence. Observations were conducted in the Johnson R-, V-, and/or B-bands, employing the standard setup for linear polarization observations (lambda/2 retarder followed by a calcite).

The data were analyzed using the semi-automatic pipeline developed at the Tuorla Observatory \citep{Hovatta2016,Nilsson2018}. This pipeline follows standard procedures, where sky-subtracted target counts were measured in the ordinary and extraordinary beams using aperture photometry (with aperture radius sizes of 1.5'' or 2.5'' depending on the seeing). Normalized Stokes parameters, polarization degree ($\Pi$), and polarization angle ($\psi$) were calculated from intensity ratios of the two beams using standard formulas \citep[e.g.][]{landi07}.

Given the low level of polarization (see below), we conducted a set of simulations to validate our error estimates and identify any potential systematic effects that may not have been considered in our analysis chain. A detailed description of this process can be found in Appendix \ref{sec:Appendix_pol}. 

In addition to our NOT observations, we supplemented our dataset with data from the Multicolour OPTimised Optical Polarimeter (MOPTOP), mounted at the 2-m Liverpool Telescope. MOPTOP features a dual-beam polarimeter with a continuously rotating half-wave plate, enabling quasi-simultaneous B-, V-, and R-band observations with a 7$\times$7 arcsec$^2$ field of view \citep{Shrestha2020}. MOPTOP observations were conducted from MJD~59816.9 to MJD~59868.8.

To enhance measurement accuracy, we combined adjacent observations by calculating a weighted average (with uncertainty) in Stokes-q and -u space. Subsequently, we converted these values to polarization degree and angle. If a $>$3$\sigma$ detection was not achieved, we increased the number of combined data points. This process allowed us to establish more robust 99.7\% ($3\sigma$) confidence interval (C.I.) upper limits overall. 

The detected values were then corrected for the depolarizing effect of the host-galaxy light \citep{andruchow08,liodakis23}. The correction depends merely on the ratio of the host galaxy flux to total flux within the aperture. To estimate the brightness of the host, we converted the Panoramic Survey Telescope \& Rapid Response System (Pan-STARRS) catalogue (DR2) mean PSF magnitudes of the galaxy to the Johnson B-, V-, and R-band magnitudes using the formulae from \citet{tonry12}, determining pre-event mean values of 20.29$\pm$0.04 mag, 19.33$\pm$0.02 mag, and 18.78$\pm$0.02 mag, respectively. The PSF magnitudes were selected, as the aperture radii used for NOT and MOPTOP observations were small; $\sim$1''--2'', that do not include the whole galaxy. In this case, we observed only a slightly higher intrinsic polarization degree compared to the observed values.

We also estimated the effect of strong Balmer emission lines in the Johnson B- and V-filters on the measured polarization degree using the NOT/ALFOSC spectrum described above. The line fluxes of H$\gamma$ and H$\beta$ are approximately 3\% and 8\% of the total flux, respectively, in the B- and V-bands. Thus, they present only a minor modification to the obtained polarization parameters, which we neglect in this paper.    

\subsection{Optical/UV photometry}

We analyzed 61 \textit{Swift} UltraViolet and Optical Telescope (UVOT) observations taken during the event between MJD 59768 and MJD 60355. All UVOT filters were utilized only during the peak emission phase, with a predominant use of the UV filters (W1/M2/W2). Source count rates, AB magnitudes, and fluxes were measured using the \textsc{uvotsource v4.3} task, which is part of the High Energy Astrophysics Software (HEASoft). When more than one exposure was taken during the telescope pointing, we combined the images using the \textsc{uvotimsum} task. Additionally, we ensured that the measured count rates were not affected by the source falling in a low-throughput area of the detector.

We also used ZTF g- and r-band data from the transient broker LASAIR \citep{smith19}.

\subsection{X-ray observations}

\subsubsection{\textit{Swift}/XRT}

We conducted a search for X-ray sources in the vicinity of AT 2022fpx by stacking 61 archival \textit{Swift} X-ray Telescope (XRT) photon counting mode images from the same set of \textit{Swift} observations mentioned above. This stacking procedure was performed using \textsc{xselect v2.5} and \textsc{ximage v4.5.1}. A 14-$\sigma$ source was identified at the coordinates RA 15:31:03.77, DEC +53:24:17.1 (position error of 3.7 arcsec), matching the location of AT 2022fpx. The mean source count rate of this source is 0.0031$\pm$0.0002 counts/s in the 0.3–1 keV band (corrected for the point spread function, sampling dead time, and vignetting). No higher energy photons were detected from the source. We further iteratively adjusted the time ranges of the stacking procedure for progressively shorter times maintaining the source at the location of AT 2022fpx with a signal-to-noise ratio larger than two at a minimum. This resulted in 24 epochs, with count rates ranging from $\sim$0.001 counts/s to $\sim$0.015 counts/s.

We also extracted a spectrum using all the XRT observations from MJD 59768 to MJD 60355, resulting in a total exposure of 124 ksec and a total of 371$\pm$25 source counts from AT 2022fpx. The spectral bins in the 0.3 to 1 keV range were grouped to a minimum signal-to-noise ratio of three, resulting in 19 spectral bins.

\subsubsection{\textit{XMM-Newton}}

We observed the source with \textit{XMM-Newton} on May 7th, 2023 (MJD 60071), 287 days after the optical peak emission. Unfortunately, the pointing was heavily affected by soft proton radiation from Earth's magnetosphere, rendering the EPIC-pn camera data unusable. Nevertheless, we were able to utilize the EPIC-MOS camera data to some extent. We extracted the source products using \textsc{SAS} version 20.0.0 following standard procedures. Background flares were filtered by excluding times when the 10$-$12 keV count rate exceeded 0.35 counts/s for MOS. This resulted in 8.5 ksec of usable exposure time. A source at the location of AT 2022fpx was detected with a significance of $\sim$14-$\sigma$, exhibiting a countrate of 0.032$\pm$0.002 counts/s and a total of 618 source counts. We further combined the MOS spectra using the \textsc{epicspeccombine} routine in \textsc{SAS}. For spectral fitting, we considered the same spectral range as for \textit{Swift}/XRT data, i.e., from 0.3 to 1 keV, grouping it similarly to a minimum S/N=3, resulting in 28 spectral bins. We performed spectral fitting using the Interactive Spectral Interpretation System \citep[\textsc{ISIS};][]{houck02} and estimated the errors on the best-fit parameter values via Monte Carlo analysis. In the modeling, we added a constant factor to account for the flux difference between the two X-ray detectors used.

\section{Results} \label{sec:results}

\subsection{Spectral reclassification of AT 2022fpx as an extreme coronal line emitter} \label{sec:classify}

\begin{figure}
    \includegraphics[width=0.5\textwidth]{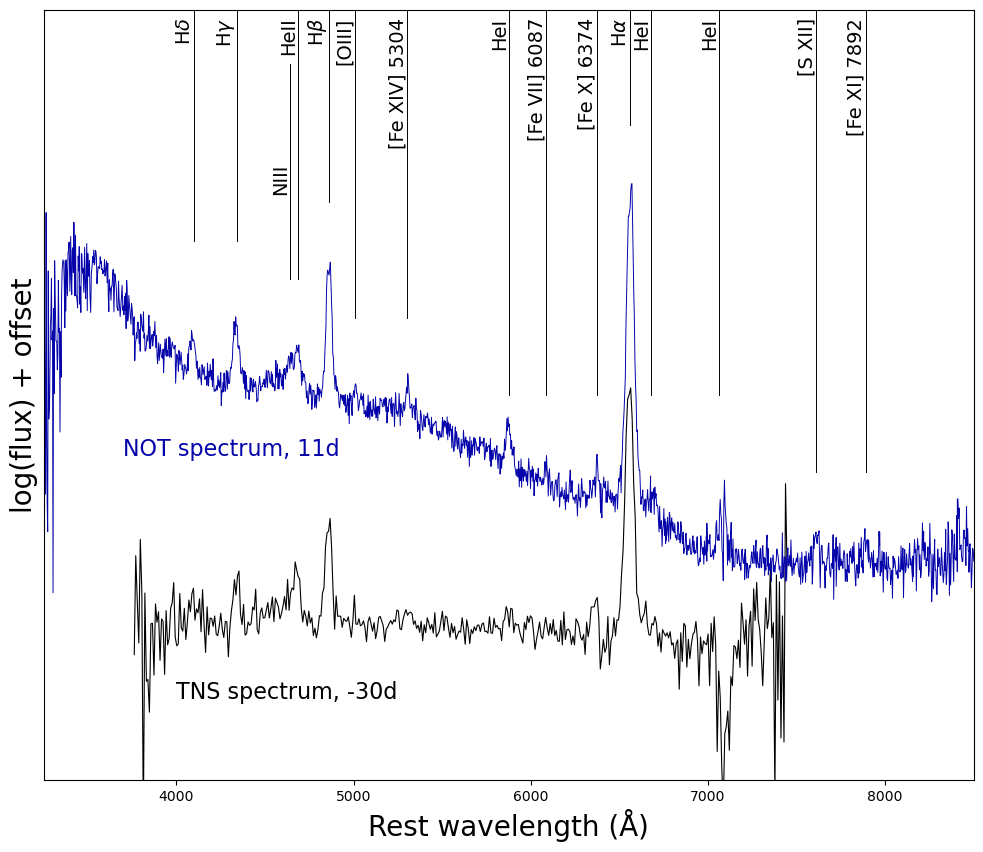}
    \caption{The TNS classification spectrum (black line) and our NOT/ALFOSC spectrum (blue line) of AT 2022fpx taken during the optical rise and close to the peak of the event, respectively. Both spectra exhibit prominent hydrogen and ionized helium lines. The NOT spectrum, acquired 41 days later than the TNS spectrum, shows a bluer continuum, featuring combined N {\sc III} and He {\sc II} line at 4635-4685 \AA, a highly-ionized silicon line at 7613 \AA\/ ([S {\sc XII}]) and highly-ionized iron lines at 5303 \AA\/ ([Fe {\sc XIV}]), 6374 \AA\/ ([Fe {\sc X}]), and 7892 \AA\/ ([Fe {\sc XI}]). Phases listed are relative to peak.}
    \label{fig:spectrum}
\end{figure}

Fig. \ref{fig:spectrum} displays the spectrum of AT 2022fpx from our NOT/ALFOSC observations. The TNS spectrum was taken 85.5 days after the initial detection, while the NOT observation occurred 126.5 days after the discovery. The line strengths and widths of the hydrogen, He I, and He II lines appear consistent in both spectra. Additionally, the NOT spectrum reveals emission lines from Bowen fluorescence (N {\sc III}) and high-ionization state `coronal' lines (S {\sc VII}, and [Fe {\sc X}] -- [Fe {\sc XIV}]), although the [Fe {\sc X}] emission line can be also seen in the TNS spectrum as well. Thus, these suggest that AT 2022fpx falls into the classification of both Bowen fluorescence TDEs and ECLEs. Previous studies, such as those by \citet{wang12} and \citet{callow24}, have employed a criterion wherein the flux of at least one coronal line exceeds [O {\sc III}] $\lambda$5007 line flux by 20\% in order to classify the source as an ECLE. In our NOT spectrum, the [Fe {\sc XIV}] $\lambda$5304 line flux surpasses this criterion, being a factor of 1.7 higher that of the [O  {\sc III}] line. Upon fitting a single Gaussian profile to the H$\alpha$ and H$\beta$ emission lines, we obtained full width at half maximum (FWHM) values of 1800$\pm 31$ and 2100$\pm 60$ km/s respectively. Conversely, fitting the He {\sc II} and N {\sc III} line blend with two Gaussian profiles yielded FWHM values of 2930$\pm 70$ km/s for He {\sc II} and 2500$\pm 100$ km/s for N {\sc III}. As discussed in Section \ref{sec:fpx}, these line widths are notably narrower than those typically observed for TDEs at early times. However, it is worth noting that TDEs can exhibit similarly narrow line widths at comparable epochs. We have obtained additional comprehensive spectroscopic follow-up of AT 2022fpx, which will be presented in a future publication (Reynolds et al., in prep).

\subsection{Host galaxy properties} \label{sec:host}

The AllWISE colors of the host galaxy (W1 - W2 = 0.147$\pm$0.063, W2 - W3 = 2.52$\pm$0.22) do not support AGN host based on the classifications of \citet{jarrett11,stern12,mateos12}. However, more recently, the infrared colors have been found not to be very accurate classifiers \citep[e.g.][]{radcliffe21}, and combining the infrared colors with optical photometry may provide more accurate results \citep{daoutis23}. Using their method, we estimate the following probabilities for the host galaxy: 53.8\% for AGN, 30.3\% for a `composite' \citep[e.g.,][]{ho97}, 8.2\% for LINER, 4.3\% for a passive, and 3.4\% for a star-forming host galaxy. The results imply a high probability of the host galaxy being an AGN or composite and quite small for being purely star-forming or passive. This is probably because the galaxy appears very red in the optical (passive) but also red in the infrared (active), which is contradicting for a purely star-forming or passive galaxy. We note that the accuracy for obtaining a correct classification for AGN is nevertheless quite low, at 56\% \citep{daoutis23}. Based on the WISE infrared colors of a sample of E+A galaxies, the rising infrared emission could be explained by the presence of post-AGB stars or a low-luminosity and/or obscured AGN \citep{alatalo17}. However, the host galaxy of AT 2022fpx is not detected in the lowest frequency WISE band (W4; 22 $\mu$m). Therefore, while we cannot completely rule out AGN host, it is safe to assume that in the case of AGN host, the nucleus has a low luminosity and it is not radiatively dominant in the system \citep{alatalo17}.

In the absence of spectra predating the transient detection, we employed SED fitting using the updated version \citep[][]{2022ApJ...927..192Y} of the Code Investigating GALaxy Emission \citep[\textsc{CIGALE};\footnote{https://cigale.lam.fr}][]{burgarella05,noll09,boquien19} to estimate the properties of the host galaxy.  Photometric data of the host galaxy were gathered from archival observations to construct its SED (see details in Appendix \ref{sec:Appendix_host}).

The SED fitting in CIGALE utilized modules for various components, including the stellar initial mass function (\texttt{bc03}), delayed star formation history (\texttt{sfhdelayed}), attenuation model (\texttt{dustatt\_modified\_starburst}), redshift (\texttt{redshifting}), AGN model (\texttt{dale2014}), and nebular emission (\texttt{nebular}), covering an extensive range of the parameter space (Table \ref{tab:pcigale_ini}). Additionally, we incorporated an additional 10\% error in the source fluxes to account for any photometric scatter, such as variations due to atmospheric conditions or low-scale variability. While it is not anticipated that the host galaxy is radiatively dominated by an AGN, we considered the potential contribution of an AGN to its SED.

\begin{figure}
    \includegraphics[width=1.0\columnwidth]{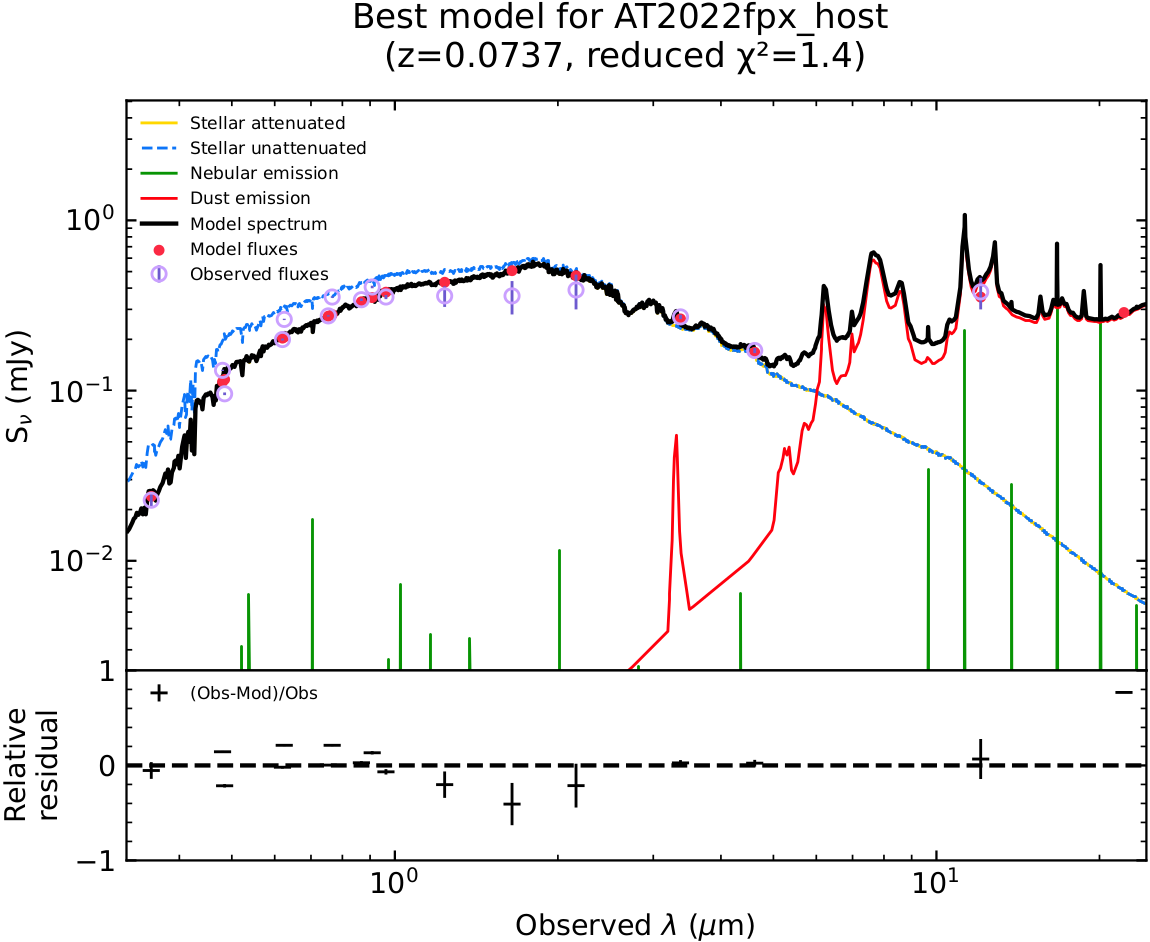}
    \caption{Best fitting model for the SED of the quiescent host galaxy of AT~2022fpx obtained by CIGALE. See text for details.}
    \label{fig:cigale}
\end{figure}

The best-fitting model SED exhibited a satisfactory fit with a low reduced chi-square value ($\chi^2_\nu = 1.4$; Fig. \ref{fig:cigale}). However, the estimations of galaxy parameters were not solely based on the best-fit model but rather on the parameter's $\chi^{2}$-weighted probability density functions (PDFs). These PDFs encompass all CIGALE models generated during the fitting process, with their weights determined by their respective $\chi^2$, reflecting how effectively each model reproduces the observed SED. This approach enables us to quantify the extent of parameter estimation variability across different models and derive robust estimations of their uncertainties. Given that the PDFs in our case do not exhibit a Gaussian distribution, the estimations rely on the mode and 68\%/32\% confidence intervals.
The $\chi^{2}$-weighted PDFs for parameters such as star formation rate (SFR), stellar mass ($M_\star$), and AGN fraction ($f_{\rm AGN}$) are provided in Fig. \ref{fig:Mstar_SFR_PDFs}.

The galaxy SED is dominated by an old stellar population with an average age of $4.7^{+0.5}_{-0.3}$~Gyr, with a more recent star formation burst with an estimated age of $1.1 \pm 0.1$~Gyr. The resulting galactic stellar mass value is $M_{\rm gal}= 7.94 \pm 0.75 \times 10^{9} \, M_\odot$. The current SFR indicated by the model is negligible, at log [SFR/($M_\odot$ yr$^{-1}$)] = $-14.1^{+0.0}_{-0.2}$, but non-negligible dust emission is required in the SED fit ($E_{\rm B-V}$ = 0.4 mag), indicating dust heated most probably by post-AGB stars. The AGN fraction is $f_{\rm AGN} = 0.0^{+0.12}_{-0.0}$ which is in agreement with other galaxy classification methods expecting no AGN contribution in the galaxy's emission. These properties could be explained if the galaxy underwent a merger with a dwarf star-forming galaxy which align well with typical E+A galaxies that are over-represented in TDE hosts \citep{arcavi14,french16,lawsmith17,hammerstein21}.

\subsection{The multiwavelength outburst evolution of AT 2022fpx}

\begin{figure*}
    \includegraphics[width=1.0\textwidth]{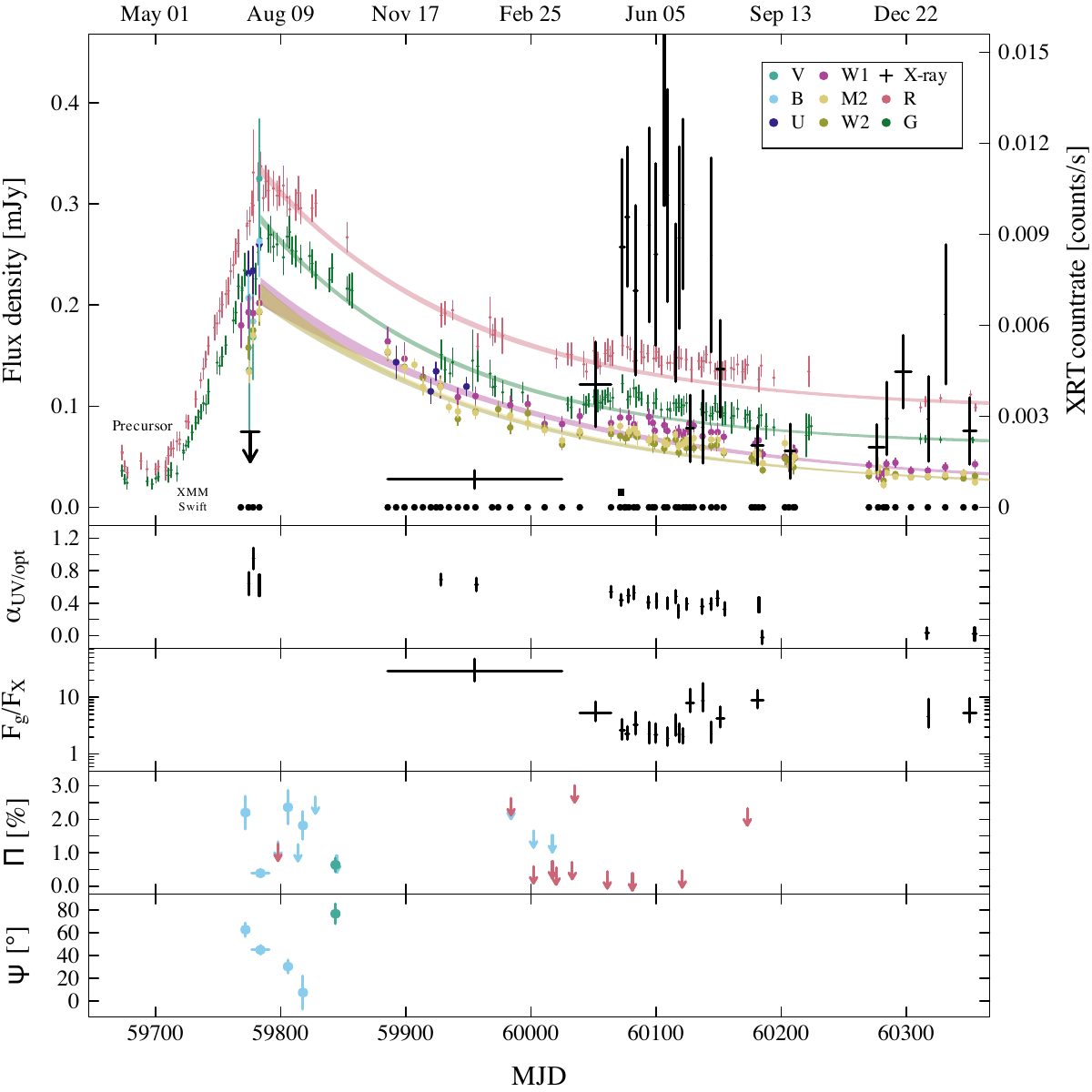}
    \caption{The multiwavelength observations of AT 2022fpx. In the top panel, colored points depict \textit{Swift}/UVOT and ZTF reddened fluxes, while black crosses represent the \textit{Swift}/XRT count rate. The black square and dots at the bottom of the panel indicate the observation times for \textit{XMM-Newton} and \textit{Swift}. Exponential decay profiles fitted to the optical/UV data are overlaid (colored lines corresponding to different wavebands), excluding the data during the high X-ray emission phase where additional flux above the exponential decay is clearly seen. In the four bottom panels, we display the spectral index of the power law model fitted to the quasi-simultaneous UV/optical SEDs (\textit{top panel}), the optical (g-band) and X-ray flux ratio (\textit{middle-top panel}), the optical polarization degree (\textit{middle-bottom panel}), and the optical polarization angle (\textit{bottom panel}) from our NOT/Liverpool Telescope observations. Different optical bands are represented by colored filled circles, as described in the figure legend in the top panel. Downward-pointing arrows indicate 3$\sigma$ upper limits (99.7\% C.I.), and the error bars show 1$\sigma$ (68\% C.I.) uncertainty.}
    \label{fig:alldata}
\end{figure*}

Fig. \ref{fig:alldata} displays the multiwavelength dataset described above for AT 2022fpx, covering the transient event from the optical rise in May 2022 to the end of 2023. We note that AT 2022fpx has not yet returned to pre-flare flux levels, and the event is still ongoing. In the top panel, we present the ZTF g- and r-band light curves, along with the \textit{Swift}/UVOT light curves and the \textit{Swift}/XRT count rate, showing variable emission. The decaying optical/UV fluxes were fitted with an exponential function in all filters. This fitting was executed using  UltraNest\footnote{\url{https://johannesbuchner.github.io/UltraNest/}} \citep{buchner21}, which utilizes the nested sampling Monte Carlo algorithm MLFriends \citep{buchner14, buchner19} to derive posterior probability distributions and Bayesian evidence. In Fig. \ref{fig:alldata} (top panel), we display the best-fit exponential functions, $F \propto \rm{exp}(-t/b)$, with decay timescales $b$ = 181$\pm$17 days (r-band), $b$ = 155$\pm$12 days (g-band), $b$ = 248$\pm$30 days (UVOT W1-band), $b$ = 202$\pm$20 days (UVOT W2-band), and $b$ = 213$\pm$18 days (UVOT M2-band). These timescales are more than two times longer than typical optical TDE decay rates \citep[e.g.,][]{liodakis23}.

Another interesting aspect of the optical light curve is a precursor event before the rise to the peak (see Fig. \ref{fig:alldata}). Precursor events in TDEs have not been commonly observed, with the exception of AT 2020wey \citep{charalampopoulos23}, where a flattening of the rise was observed for approximately eight days, and AT 2019azh \citep{faris23} with a similar flattening from 10 to 30 days before the light curve peak. Similar to the slow timescale of the decay, the precursor event in AT 2022fpx is taking place much earlier at $\sim$70 days before the light curve peak. 
 
We also fitted the six-band \textit{Swift}/UVOT SEDs, or in cases where only UV bands are available, the quasi-simultaneous ($\pm$2 days) UVOT + ZTF SEDs, with black body and power law models. In general, the power law model fits the SEDs better. In Fig. \ref{fig:alldata} (2nd panel), we show the spectral index of the power law model, $F_\nu \propto \nu^{-\alpha_{\rm UV/opt}}$, which decreases steadily over the decay from $\alpha_{\rm UV/opt}\sim1$ to $\alpha_{\rm UV/opt}\sim0$, indicating spectral flattening. A black body model at the peak gives a rest-frame bolometric luminosity of $L_{\rm bol, peak} = 5.9 \pm 1.1 \times 10^{43}$ erg/s and a temperature of $T_{\rm peak} = 16600\pm850$ K.

We also calculated the optical-to-X-ray flux ratio ($F_{\rm g}/F_{\rm X}$) by converting the \textit{Swift}/XRT count rate to X-ray flux in the 0.1--1 keV band using the best-fit X-ray model fitted to the \textit{XMM-Newton} and \textit{Swift}/XRT data (see Section \ref{sec:lateX}). The flux ratio evolution plotted in Fig. \ref{fig:alldata} (3rd panel) shows that the ratio reached a value of about 2 at the phase of peak X-ray emission, which is a typical ratio for an optical TDE with late-time X-ray brightening \citep{gezari17}.

\subsection{Optical polarization results}

The bottom two panels in Fig. \ref{fig:alldata} shows the polarization results from our NOT and Liverpool Telescope observations. The source consistently exhibits a low polarization degree ($\Pi\lesssim3$\%) throughout our observations. During the peak optical emission, we successfully detected polarization in the B-band in both ALFOSC and MOPTOP observations and additionally in the V-band for ALFOSC. At later times, we constrain the polarization degree to $\Pi<0.5\%$ at the 99.7\% C.I. (R-band), suggesting that the source became unpolarized. The polarization degree in the B-band detections varied from $\Pi\sim$0.4\% to $\Pi\sim$2.7\%, while for the V-band detection it was $\Pi\sim0.8\%$.

We observed significant variability in both the polarization degree and angle from AT 2022fpx. Previous optical polarization observations of TDEs have typically shown polarized emission after the peak with decreasing polarization degree as the event progresses \cite[e.g.,][]{Lee2020, liodakis23}. Although the overall trend seems to match previous results, this is the first time we observe short-term variability. The first B-band polarization detection ($\Pi\sim2.5\%$) takes place just before the peak of the optical light curve, the second B-band observation ($\Pi\sim0.4\%$) at the peak, and the remaining detections in B- and V-bands occur after the peak ($\Pi=0.8\%-2.7\%$). At the same time, we observe a change in the B-band polarization angle from $\psi=62^\circ$ to $\psi=7^\circ$ in the form of a continuous smooth rotation. The V-band detection yields a polarization angle  $\psi=76^\circ$, consistent within uncertainties with the first B-band detection, but not part of the B-band rotation trend.

After the decay has progressed more than 200 days, we observe a brightening in X-rays (see details in Section \ref{sec:lateX}). The polarization degree in the R-band remains undetected at the $3\sigma$ level. Lowering the detection threshold to 2.6$\sigma$ (99\% C.I.), we find two `significant detections' with polarization degree around $\Pi\sim0.4\%-0.7\%$ (from MJD 60061 to MJD 60081). Interestingly, these `detections' occur around the time the source starts to brighten in X-rays, after which the polarization degree drops to $\Pi<0.4\%$ (99\% C.I.). We discuss further the possible scenarios for the polarization variability in Section \ref{sec:pol_discuss}.

\subsection{Late-time X-ray brightening} \label{sec:lateX}

\begin{figure}
    \includegraphics[width=1.0\columnwidth]{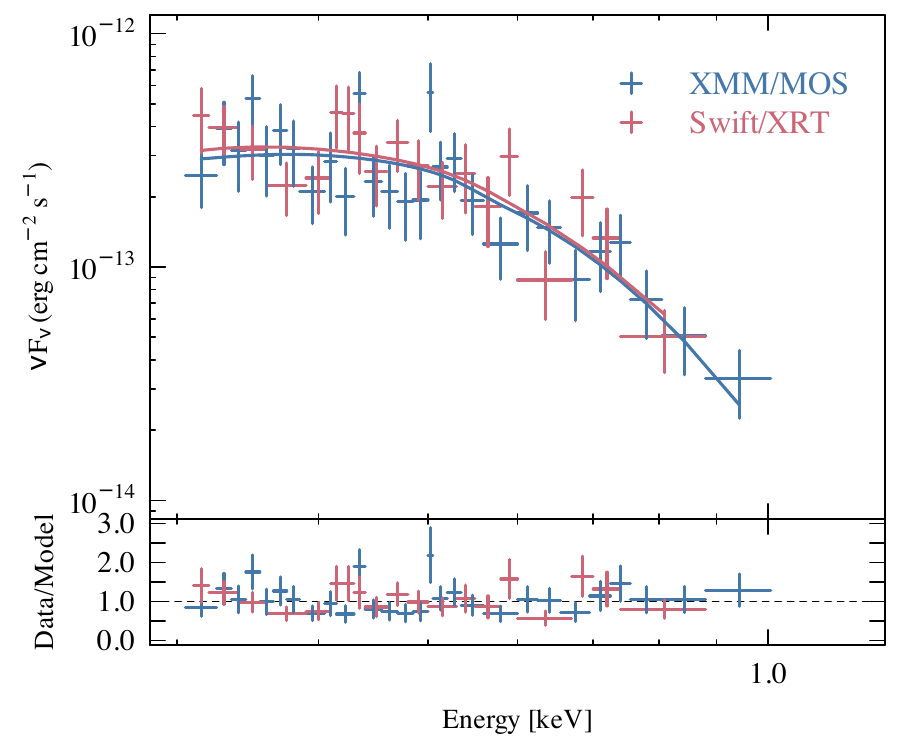}
    \caption{The time-averaged (\textit{Swift}/XRT) and snapshot (\textit{XMM}/MOS) X-ray spectra of AT 2022fpx fitted with the same absorbed black body model. XRT spectrum is normalized to match the MOS spectrum.}
    \label{fig:xspec}
\end{figure}

The X-rays started to rise around MJD 60050 with possible low X-ray activity before that at a level of $(9\pm3)\times10^{-4}$ counts/s (Fig. \ref{fig:alldata}), corresponding to an unabsorbed flux of $(2.8\pm0.9)\times10^{-14}$ erg s$^{-1}$ cm$^{-2}$ in the 0.3--1.0 keV band. This assumes a black body spectrum with a temperature of 95 eV and a hydrogen column density of 1.25$\times$10$^{20}$ atoms cm$^{-2}$ along the line of sight (see below; Fig. \ref{fig:xspec}). The rise in X-rays appear to occur simultaneously with the small rise in UV/optical fluxes during the decay, as indicated by their deviation from the exponential decay profile in Fig. \ref{fig:alldata}.  At the UV/optical peak, the X-rays are not detected, with a 3$\sigma$ upper limit of $2.5\times10^{-3}$ counts/s ($7.7\times10^{-14}$ erg s$^{-1}$ cm$^{-2}$). The X-ray emission lasted about 100 days at a more or less constant level, but dimmed down after that. However, recent observations show that the X-ray emission is recovering again, possibly indicating the start of another `flare' or a periodic behavior.

We fitted both the \textit{XMM-Newton}/MOS and the time-integrated \textit{Swift}/XRT spectra with an absorbed and redshifted black body model using Cash statistics and subplex optimization in ISIS. The agreement of the time-integrated \textit{Swift}/XRT spectrum spanning $\sim$400 days and the \textit{XMM-Newton} snapshot spectrum indicates that the shape of the spectrum remains constant throughout the X-ray brightening (Fig. \ref{fig:xspec}). We fixed the interstellar absorption at $N_{\rm{H}}=1.25 \times 10^{20}$ atoms/cm$^{2}$ \citep{HI4PI} by adding a \textsc{tbabs} component to the model, using abundances from \citet{wilms00}. The best-fit model has Cash statistics of 49.5 with 44 degrees of freedom and indicates a temperature of 95$\pm5$ eV. The rest-frame bolometric black body luminosity during the \textit{XMM-Newton} observation at the source distance of 311 Mpc is (8.7$\pm$1.2)$\times 10^{42}$ erg s$^{-1}$. While the temperature is a reasonable value for a TDE \citep[e.g.][]{kajava20}, the luminosity is relatively low, corresponding to the Eddington luminosity of a $\sim 7 \times 10^{4} \, \rm{M}{\odot}$ black hole or an Eddington fraction of 0.007 for a $10^{7} \, \rm{M}{\odot}$ black hole.

\subsection{Black hole mass estimates}

We estimated the mass of the black hole through various methods, including fitting the observed light curves with The Modular Open Source Fitter for Transients \citep[\textsc{MOSFiT;}\footnote{\url{https://github.com/guillochon/MOSFiT}}][]{guillochon18} resulting in $\rm{log} \, M_{\rm{BH}}/M_\odot = 7.46^{+0.12}_{-0.05}$, and using the code \textsc{TDEmass}\footnote{\url{https://github.com/taehoryu/TDEmass}} \citep{ryu20}, yielding $\rm{log} \, M_{\rm{BH}}/M_\odot = 6.71^{+0.15}_{-0.18}$. We utilized the TDE model \citep{mockler19} within \textsc{MOSFiT} and the observed light curves in different bands, including \textit{Swift}/UVOT W1-, M2-, and W2-bands, as well as ZTF $r$- and $g$-bands, as input. The code assumes that the optical-to-UV emission is reprocessed accretion disk emission in the surrounding gas. \textsc{TDEmass} solves two non-linear algebraic equations derived in \citet{ryu20} that arise from the outer shock model, where the optical/UV emission is generated by dissipation in shocks near the debris' orbital apocenter. The parameters required for solving the equations are the observed peak UV/optical luminosity and the observed temperature at peak luminosity, which we derive from a black body model fit to the \textit{Swift}/UVOT data taken at MJD 59783 coinciding with the ZTF light curve peak. 

We also used empirical relations with the black hole mass involving peak luminosity and radiated energy \citep{mummery24}, which resulted in $\rm{log} \, M_{\rm{BH}}/M_\odot = 6.80 \pm 0.54$ and $\rm{log} \, M_{\rm{BH}}/M_\odot = 7.19 \pm 0.44$, respectively. The relation of the black hole mass with the plateau luminosity in \citet{mummery24} yielded a much higher value, $\rm{log} \, M_{\rm{BH}}/M_\odot = 8.4$, which is not surprising, as AT 2022fpx has not yet reached the plateau phase and remains quite luminous. We also utilized an empirical relation of the black hole mass with the host galaxy mass \citep{greene20, mummery24}, resulting in $\rm{log} \, M_{\rm{BH}}/M_\odot = 6.62 \pm 0.58$ using the host galaxy mass derived in Section \ref{sec:host}. If the above estimates for the black hole mass are independent, the global mean would be $\rm{log} \, M_{\rm{BH}}/M_\odot = 6.95 \pm 0.53$.

\section{Discussion} \label{sec:discuss}

\subsection{On the variable polarization} \label{sec:pol_discuss}

The low optical polarization degree observed in optical TDEs has often been interpreted as arising from scattering from free electrons in a reprocessing layer surrounding the black hole \citep{leloudas22, Charalampopoulos2023}. While our results would indeed be consistent with such a scenario, the erratic variability in the polarization degree and rotation in the polarization angle observed in AT 2022fpx cannot easily fit into the reprocessing picture. In this scenario, the X-ray brightening after the peak is attributed to the reprocessing layer becoming optically thin to X-rays. Using the global mean estimate for the black hole mass, we find the necessary time for the transition to the optically thin regime to be $t_{\text{\rm thin}}\approx14$ days \citep{metzger16}, which is $\sim$20 times shorter than the observed time lag between the optical and X-ray peak emissions.

Previously, we interpreted changes in the polarization angle as arising from the shift in shock dominance within the outer shock scenario \citep{liodakis23}. In this scenario, intersecting stellar streams near the pericenter of the stellar orbit form a nozzle shock that dissipates orbital energy \citep{bonnerot22}. However, this energy loss is likely insufficient to rapidly circularize the flow and lead to the formation of the accretion disk \citep{piran15,steinberg24}. Instead, the stellar flow collides with itself further away from the black hole, producing strong shocks that emit in the optical. According to this scenario, we expect the pericenter shock to form and dominate the emission during the rising phase of the outburst \citep{steinberg24}. The pre-peak polarized emission would then correspond to emission from the pericenter shock. The subsequent drop in the polarization degree at the optical peak could result from the formation of the outer shocks, as emission from multiple polarized components with different polarization angles could lead to depolarization \cite[e.g.,][]{Liodakis2020}. The change in the polarization angle coincides with the drop in the polarization degree. However, the polarization angle continues to change while the polarization degree recovers back to pre-peak level. After the optical peak, the polarization degree drops again, but now the polarization angle returns to the pre-peak values (although in a different optical band). This erratic variability and changes in the optical polarization angle could be interpreted as contributions from different shocks at different times. However, without dedicated simulations to evaluate the relative contribution of the different shocks to their composite emission, we cannot reach firm conclusions. Using the black hole mass estimate from \textsc{TDEmass}, we can estimate the return time of the most-bound material to be $t_0=52\pm10$ days (following \citealt{piran15}). Thus, the phase of the outer shock emission is expected to take place around $1.5-3\times t_0$ \citep{shiokawa15}, precisely the time interval of the observed polarization variability. In addition, the expected time from the disruption to the optical peak ($1.5 \times t_0$) is $\approx78$ days, matching the outburst rise time well. In this scenario, the brightening in X-rays can be interpreted as the delayed formation of the accretion disk \cite[e.g.,][]{kajava20}. The accretion disk is expected to form between $3-10 \times t_0$ \citep{shiokawa15}, i.e., $\sim$150--520 days, consistent with the observed behavior.

We observe a potential (2.6$\sigma$) increase in the polarization degree coinciding with the rise of the X-ray emission. Interestingly, the X-ray emission lasts for only $\sim$100 days at a high level, accompanied by only a mild increase in the optical flux and a flattening of the optical/UV spectrum during the same period. The observed low degree ($\Pi<0.5\%$) of polarization during the X-ray flare is typical of values found for the polarization from X-ray binary accretion disks \cite[e.g.,][]{Krawczynski2022} and AGN \cite[e.g.,][]{Marin2023}.

Another scenario that could possibly explain the observed polarization behavior is scattering from a compact, clumpy torus \citep{Marin2015,Marin2015-II}. In this scenario, the degree of optical polarization is expected to be low. However, both the polarization degree and angle depend on the inclination of the system to the line of sight which should not vary unless the accretion disk undergoes precessing motion (which could be the case if the orbital plane of the disrupted star and the spin axis of the black hole are misaligned; \citealt{Teboul2023}). In that case, the intrinsic emission changes angle to the scatterer leading to erratic variability of the polarization degree and both smooth and sudden changes of the polarization angle \cite[see Fig. 1 in][]{Marin2015-II}. The observed polarization angle rotation rate of change in the B-band is $\dot{\psi}$ = (1.2$\pm$0.3)$^\circ$/day. Assuming polarization angle variations are achromatic, the rate of change of the polarization angle from the last B-band detection to the V-band is twice as fast $\dot{\psi}$ = (2.6$\pm$0.6)$^\circ$/day, although it is still within 2$\sigma$ from the B-band rotation rate and could imply that the polarization angle rate of change remained the same. The lack of detectable polarization at late times could be attributed to the end of the disk precession. That bounds the alignment timescale of the disk to between $\sim$140--284 days. Following \cite{Teboul2023}, we can estimate the allowed parameter space for the mass and spin of the black hole for the disruption of a solar mass star. We find black hole mass values consistent with the estimates above; however, they require a fairly low black hole spin; $\alpha\sim0.1$. While the value of the spin is low, similar results were found for AT2020ocn that showed quasi-periodic X-ray eruptions attributed to Lense-Thirring precession of the accretion disk \citep{Pasham2024}. Under this scenario, the brightening of X-rays can be interpreted as the central source becoming temporarily unobscured from the clumpy torus material. This could also explain both the flare profile of the observed X-rays (and the recurrence of the X-ray emission) as well as the very low levels of polarization degree. One might expect the polarization degree to rise again after the central source becomes obscured. Unfortunately, our optical polarization observations did not continue after the first X-ray flare, so we cannot confirm the possibility of this scenario.

\subsection{The nature of AT 2022fpx}

\begin{figure}
    \includegraphics[width=1.0\columnwidth]{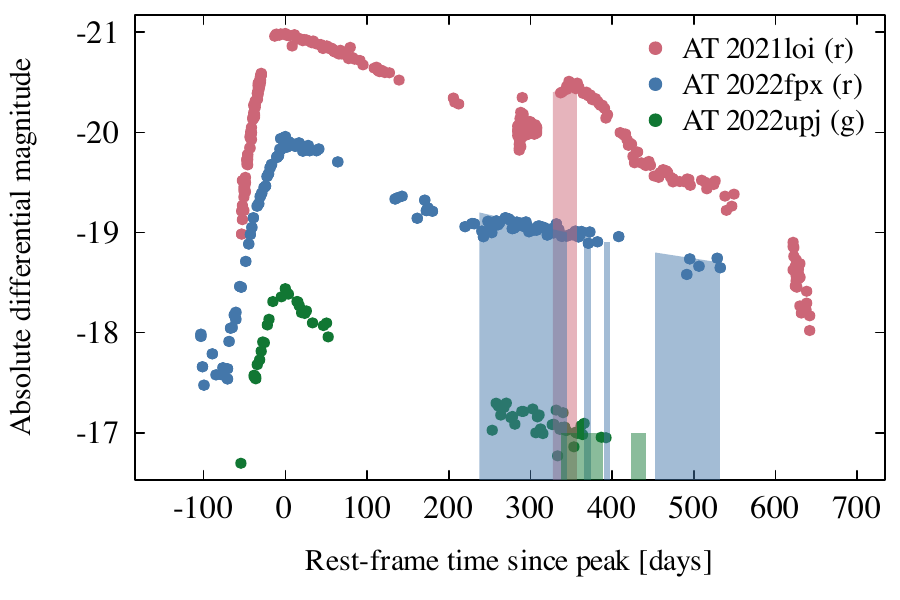}
    \caption{Comparison of the rest-frame ZTF $r$- or $g$-band light curves from AT 2021loi \citep[a candidate BFF source;][]{makrygianni23}, AT 2022fpx, and AT 2022upj (both coronal line sources). The translucent colored regions represent the times of detected X-ray emission from the sources.}
    \label{fig:compare}
\end{figure}

AT 2022fpx is among the few optical TDEs exhibiting coronal line emission during the flare decay, alongside AT 2017gge \citep{onori22}, AT 2019qiz \citep{short23}, and AT 2022upj \citep{newsome22}. The appearance of coronal lines varies from source to source, with AT 2017gge and AT 2019qiz showing the lines at later times, approximately $\sim$200 and $\sim$400 days after the optical peak, respectively, while AT 2022upj and AT 2022fpx exhibit the lines earlier, at approximately $\sim$41 and $\lesssim$86 days after the optical peak, respectively. 

Additionally, the appearance of X-rays differs; AT 2017gge \citep{onori22}, and AT 2022fpx exhibit late-time emission, whereas AT 2019qiz \citep{nicholl20} show X-ray emission already at the optical peak. For AT 2022upj, we examined the available \textit{Swift}/XRT data, and found clear detections starting from $\sim$350 days from the optical peak (see Fig. \ref{fig:compare}) aligning with the late-time detections of X-rays from AT 2017gge and AT 2022fpx. Thus, the relationship between the ionizing X-ray emission and the coronal lines appears complex. However, AT 2019qiz is one of the closest TDEs, which could explain the earlier X-ray detection.  

In addition to the coronal lines, AT 2022fpx shares several properties with BFF events, including the clear identification of Bowen lines, a slow decay of optical fluxes, and a rebrightening phase occurring approximately a year after the initial decline \citep{makrygianni23}. While for other BFFs, it has appeared clearer that the observed events could arise from some form of reactivation or flare of an already identified AGN \citep{trakhtenbrot19, makrygianni23}, in the case of AT 2022fpx, the AGN nature of the galaxy is uncertain, and several pieces of evidence point against it, including the ECLE nature (see Section \ref{sec:classify}), host galaxy SED, and the mid-infrared colors (see Section \ref{sec:host}). Therefore, AT 2022fpx could serve as a crucial source to link at least some BFFs to TDEs.

The resemblance of the optical light curve of AT 2022fpx to a well-sampled BFF, AT 2021loi \citep{makrygianni23}, is striking (Fig. \ref{fig:compare}). The light curves have similar shapes, with only the rise being faster and the rebrightening more pronounced in AT 2021loi compared to AT 2022fpx. Additionally, the timing of the rebrightening is almost identical. This similarity could suggest potentially similar black hole masses in these systems, around $M_{\rm BH} \sim 10^7 \, M_\odot$. The difference in rebrightening amplitudes might be attributed to AT 2021loi being more obscured, leading to X-rays being reprocessed to the optical regime (although not suggested by \citealt{makrygianni23} due to the prominently blue continuum of the source), while AT 2022fpx is either less obscured or has more clumpy blocking matter, allowing more X-rays to propagate without reprocessing. To explore this further, we examined the \textit{Swift}/XRT data of AT 2021loi up to date and found a 3$\sigma$ source at the correct location by stacking all available photon counting mode data. Notably, most of the X-ray counts coincide with the time of the rebrightening (Fig. \ref{fig:compare}). The combined \textit{Swift}/XRT countrate of the two pointings taken between MJD 59754 and MJD 59786 (corresponding to 327--357 days after peak) is 0.0025$\pm$0.00095 (2.6$\sigma$). Whether or not this is a true detection, it potentially indicates that the onset or increase of the accretion disk emission can occur later in BFFs, similar to other optical TDEs.

The comparable light curve profiles of AT 2022fpx and AT 2021loi argue against the origin from AGN variability, as one would expect AGN flares to be stochastic events and not follow a similar trend. This lends more weight to the TDE classification; however, the long-term light curve and the lower velocities of the Balmer lines still need clarification in this context. If some BFFs are TDEs occurring in low-luminosity or quiet AGN, our line-of-sight to the central source could be obstructed by a dusty torus, and the longer time scales might be influenced by the more complex surroundings and a potentially low-luminosity accretion flow already in place. This scenario aligns well with the observed polarization properties and the late-time X-ray emission seen in AT 2022fpx.

\section{Conclusions} \label{sec:conclude}

In this paper, we presented the optical polarization monitoring observations of AT 2022fpx together with accompanying UV/optical photo- and spectrometry, and X-ray observations. The optical spectra taken around $\sim$100 days after the optical peak shows highly-ionized (`coronal') lines making AT 2022fpx one of the few TDEs showing these lines during the optical decay. The source exhibited also peculiar, variable but low-polarized emission at the optical peak of the transient event including a clear rotation of $\sim$50 degrees of the polarization angle. In addition, X-ray emission was observed later on in the decay coinciding with a small rise in the optical/UV fluxes, but displaying more flare-like emission profile that potentially is re-emerging currently. 

We attribute the polarization behavior and late-time X-rays either to the outer shock scenario, which has been established to explain the observed behavior of optical TDEs, or perhaps more plausibly, to a clumpy torus surrounding the supermassive black hole. AT 2022fpx is the first source showing properties from both ECLEs and BFFs, characterized by coronal and Bowen lines, slowly decaying optical/UV light curve, and a rebrightening event (more prominent in X-rays than optical). The relation to possible dormant or obscured AGN remains an interesting and intriguing aspect of the BFF sources. However, the AGN nature in the case of AT 2022fpx is not clear, and the presence of coronal lines suggests a TDE classification. Thus, AT 2022fpx could be a key source in linking BFF sources and ECLEs to long-lived TDEs.

\section*{Acknowledgements}


K.I.I.K. has received funding from the European Research Council (ERC) under the European Union’s Horizon 2020 research and innovation programme (grant agreement No. 101002352, PI: M. Linares). K.I.I.K. acknowledges the financial support from the visitor and mobility program of the Finnish Centre for Astronomy with ESO (FINCA), funded by the Academy of Finland grant No. 306531. 

I.L was funded by the European Union ERC-2022-STG - BOOTES - 101076343. Views and opinions expressed are however those of the author(s) only and do not necessarily reflect those of the European Union or the European Research Council Executive Agency. Neither the European Union nor the granting authority can be held responsible for them. I.L. was supported by the NASA Postdoctoral Program at the Marshall Space Flight Center, administered by Oak Ridge Associated Universities under contract with NASA. 

E.L. was supported by Academy of Finland project Nos. 317636, 320045 and 346071. 

P.C. acknowledges support via Research Council of Finland (grant 340613)

K.K. was supported by the Czech Science Foundation project No.22-22643S.


LASAIR is supported by the UKRI Science and Technology Facilities Council and is a collaboration between the University of Edinburgh (grant ST/N002512/1) and Queen’s University Belfast (grant ST/N002520/1) within the LSST:UK Science Consortium. 

ZTF is supported by National Science Foundation grant AST-1440341 and a collaboration including Caltech, IPAC, the Weizmann Institute for Science, the Oskar Klein Center at Stockholm University, the University of Maryland, the University of Washington, Deutsches Elektronen-Synchrotron and Humboldt University, Los Alamos National Laboratories, the TANGO Consortium of Taiwan, the University of Wisconsin at Milwaukee, and Lawrence Berkeley National Laboratories. Operations are conducted by COO, IPAC, and UW. 

Based on observations made with the Nordic Optical Telescope, owned in collaboration by the University of Turku and Aarhus University, and operated jointly by Aarhus University, the University of Turku and the University of Oslo, representing Denmark, Finland and Norway, the University of Iceland and Stockholm University at the Observatorio del Roque de los Muchachos, La Palma, Spain, of the Instituto de Astrofisica de Canarias.  

This research has made use of data and/or software provided by the High Energy Astrophysics Science Archive Research Center (HEASARC), which is a service of the Astrophysics Science Division at NASA/GSFC. 

Based on observations obtained with XMM-Newton, an ESA science mission with instruments and contributions directly funded by ESA Member States and NASA. 

The Liverpool Telescope is operated on the island of La Palma by Liverpool John Moores University in the Spanish Observatorio del Roque de los Muchachos of the Instituto de Astrofisica de Canarias with financial support from the UKRI Science and Technology Facilities Council (STFC) (ST/T00147X/1)

The Pan-STARRS1 Surveys (PS1) and the PS1 public science archive have been made possible through contributions by the Institute for Astronomy, the University of Hawaii, the Pan-STARRS Project Office, the Max-Planck Society and its participating institutes, the Max Planck Institute for Astronomy, Heidelberg and the Max Planck Institute for Extraterrestrial Physics, Garching, The Johns Hopkins University, Durham University, the University of Edinburgh, the Queen's University Belfast, the Harvard-Smithsonian Center for Astrophysics, the Las Cumbres Observatory Global Telescope Network Incorporated, the National Central University of Taiwan, the Space Telescope Science Institute, the National Aeronautics and Space Administration under Grant No. NNX08AR22G issued through the Planetary Science Division of the NASA Science Mission Directorate, the National Science Foundation Grant No. AST-1238877, the University of Maryland, Eotvos Lorand University (ELTE), the Los Alamos National Laboratory, and the Gordon and Betty Moore Foundation.

This publication makes use of data products from the Two Micron All Sky Survey, which is a joint project of the University of Massachusetts and the Infrared Processing and Analysis Center/California Institute of Technology, funded by the National Aeronautics and Space Administration and the National Science Foundation.

Funding for the Sloan Digital Sky Survey IV has been provided by the Alfred P. Sloan Foundation, the U.S. Department of Energy Office of Science, and the Participating Institutions. SDSS-IV acknowledges support and resources from the Center for High Performance Computing  at the University of Utah. SDSS-IV is managed by the Astrophysical Research Consortium for the Participating Institutions of the SDSS Collaboration including the Brazilian Participation Group, the Carnegie Institution for Science, Carnegie Mellon University, Center for Astrophysics | Harvard \& Smithsonian, the Chilean Participation Group, the French Participation Group, Instituto de Astrof\'isica de Canarias, The Johns Hopkins University, Kavli Institute for the Physics and Mathematics of the Universe (IPMU) / University of Tokyo, the Korean Participation Group, Lawrence Berkeley National Laboratory, Leibniz Institut f\"ur Astrophysik Potsdam (AIP),  Max-Planck-Institut f\"ur Astronomie (MPIA Heidelberg), Max-Planck-Institut f\"ur Astrophysik (MPA Garching), Max-Planck-Institut f\"ur Extraterrestrische Physik (MPE), National Astronomical Observatories of China, New Mexico State University, New York University, University of Notre Dame, Observat\'ario Nacional / MCTI, The Ohio State University, Pennsylvania State University, Shanghai Astronomical Observatory, United Kingdom Participation Group, Universidad Nacional Aut\'onoma de M\'exico, University of Arizona, University of Colorado Boulder, University of Oxford, University of Portsmouth, University of Utah, University of Virginia, University of Washington, University of Wisconsin, Vanderbilt University, and Yale University.

This publication makes use of data products from the Wide-field Infrared Survey Explorer, which is a joint project of the University of California, Los Angeles, and the Jet Propulsion Laboratory/California Institute of Technology, funded by the National Aeronautics and Space Administration.

This work was supported by the MINECO (Spanish Ministry of Economy, Industry and Competitiveness) through grant ESP2016-80079-C2-1-R (MINECO/FEDER, UE) and by the Spanish MICIN/AEI/10.13039/501100011033 and by "ERDF A way of making Europe" by the “European Union” through grants RTI2018-095076-B-C21 and PID2021-122842OB-C21, and the Institute of Cosmos Sciences University of Barcelona (ICCUB, Unidad de Excelencia ’Mar\'{\i}a de Maeztu’) through grant CEX2019-000918-M.

\section*{Data Availability}

The Liverpool and NOT data are available on demand. ZTF lightcurves are available at the ZTF community brokers' (e.g., LASAIR, ALeRCE) websites. \textit{XMM-Newton}/\textit{Swift} data is available at HEASARC.



\bibliographystyle{mnras}
\bibliography{bibliography} 




\appendix



\section{Error estimates and systematic effects of the NOT polarization data}
\label{sec:Appendix_pol}

In principle, the standard dual-beam sequence of the polarization data reduction pipeline should cancel out any effects caused by the gain difference between the ordinary and extraordinary beams. However, some of our observations were made during a full moon, resulting in a relatively high and sometimes variable background. Specifically, two epochs were conducted under bright sky conditions, where the sky background was approximately 10 times brighter than in moonless conditions. Additionally, three epochs occurred under grey sky conditions, with the sky background being roughly three times brighter, while the remaining epochs were conducted under dark conditions. Typically, the background exhibited changes of 5--20\% during the image sequence in dark and grey conditions, except for one observing epoch where the rising Moon caused the sky background to increase by a factor of $\sim$3. During the two observations conducted under bright conditions, the sky level changed by less than 20\%.

Our flat-fielding procedure does not completely eliminate dust features, which could potentially affect the measured polarization properties during bright sky conditions. Therefore, we generated mock images of the source with zero polarization degree for those epochs where significant polarization was detected. We analyzed these mock images similarly to the observed data. The mock data consisted of images whose properties, such as the number, size, noise parameters, background level, rate of background level change, and signal-to-noise ratio per beam, accurately corresponded to the conditions of a given observing epoch. The mock images were also multiplied by an image containing the residual dust features.

\begin{figure}
    \includegraphics[width=0.48\textwidth]{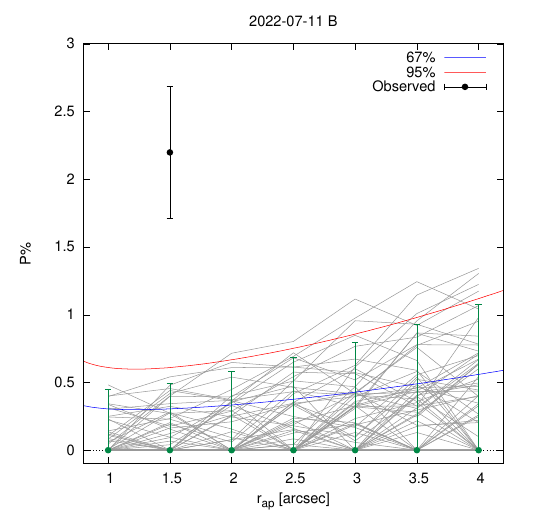}
    \caption{An example of the usage of mock data to validate the accuracy of our error determination. The grey lines represent growth curves (the degree of polarization, $P$, as a function of the aperture radius) for a  hundred mock data sets corresponding to a source with zero polarization. The green symbols depict one such curve, accompanied by 2$\sigma$ error bars derived from our analysis chain. The blue and red lines indicate the 67\% and 99\% confidence limits, respectively, using the standard formula (see text for details). The black symbol denotes the polarization degree derived from the actual data.}
    \label{fig:polsimu}
\end{figure}

We created one hundred sets of mock images per epoch and computed the polarization degree of the source based on the measurement aperture radius. The distribution of the polarization degree was then compared to the magnitude of the errors returned by our analysis chain at each aperture radius. Figure \ref{fig:polsimu} displays the results for the first night of our observing campaign, during which the sky background was particularly high. Similar results were obtained for the other epochs; specifically, the distribution of polarization degree derived from the mock data (grey lines in Fig. \ref{fig:polsimu}) aligns with the errors derived by our analysis chain and with the standard formula $\sigma_P = 0.5$/SNR, where $\sigma_p$ is the error of the polarization degree, and SNR is the signal-to-noise ratio of a single retarder position image. Our simulations demonstrate that the bright and varying background, as well as the flat-field residuals, do no significantly impact our measurements. While there is some indication that errors are slightly underestimated at larger apertures, it is worth noting that our measurements were conducted using apertures smaller than the affected region.

\section{Host galaxy SED and determination of its potential variability}
\label{sec:Appendix_host}

\begin{figure}
    \includegraphics[width=0.48\textwidth]{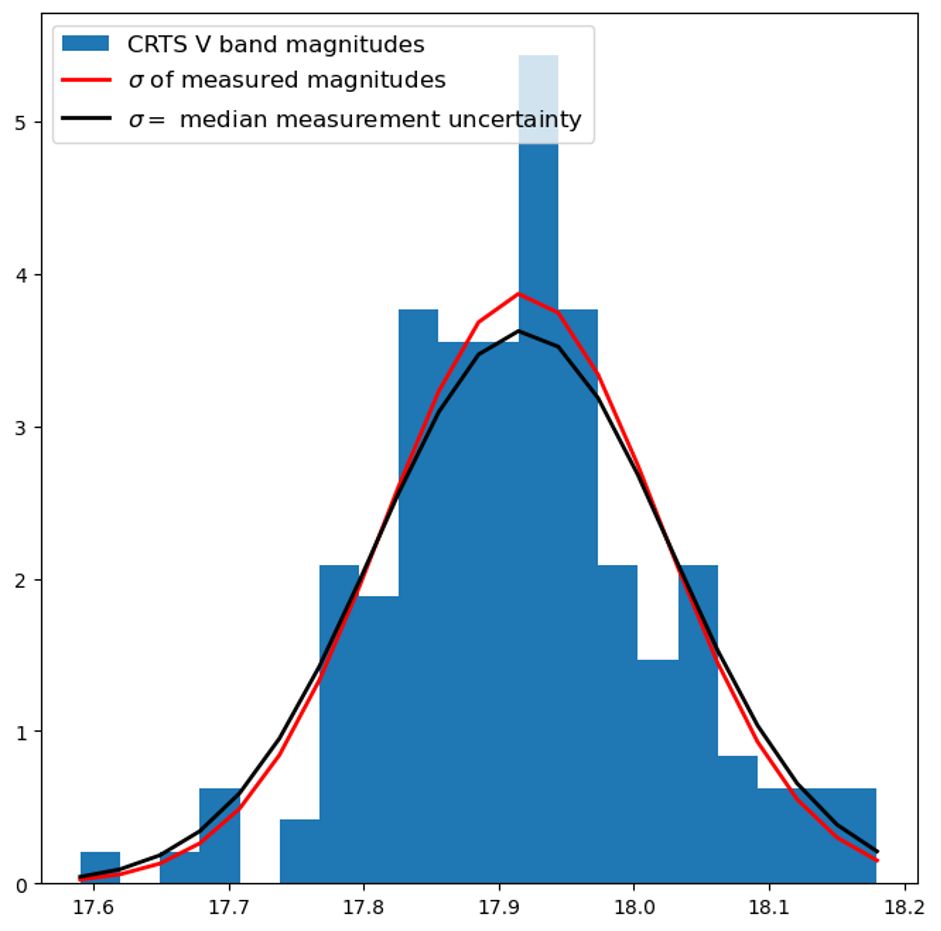}
    \caption{A magnitude histogram in V-band of the host galaxy from CRTS spanning seven years of observations from 2006 to 2013. The red line depicts the expected distribution of the measured magnitudes under the assumption of standard error, while the black line represents the expected distribution assuming a median photometric uncertainty of the individual measurements. All the measured magnitudes are consistent with a constant V-band magnitude of 17.9 mag.}
    \label{fig:CRTS}
\end{figure}

\begin{figure}
    \includegraphics[width=0.48\textwidth]{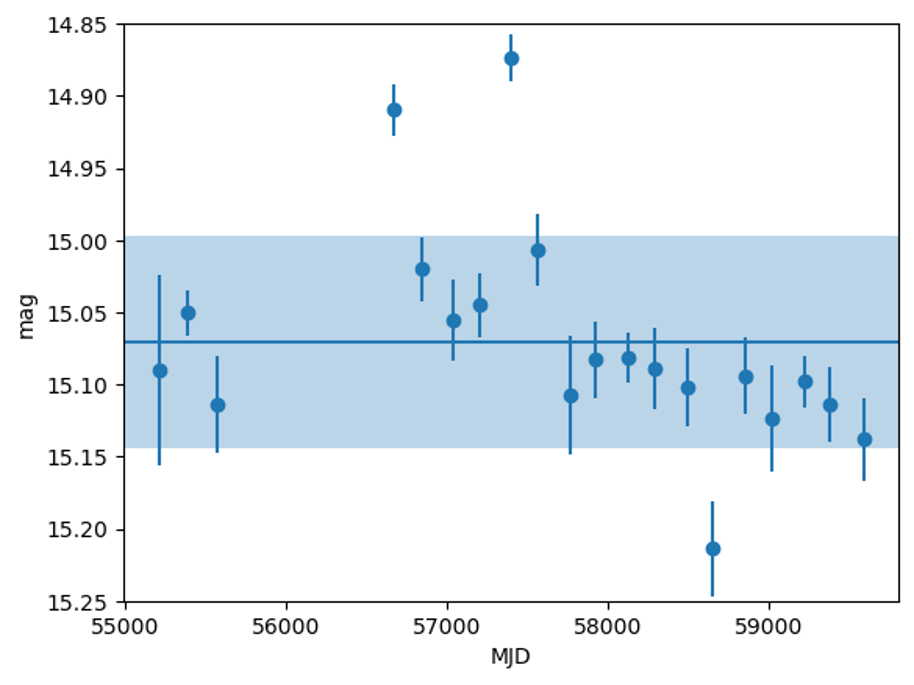}
    \caption{Time evolution of the host galaxy in W1-band (3.4 micron) from NEOWISE survey. The shaded region show the standard deviation of the data.}
    \label{fig:NEOWISE}
\end{figure}

We compiled the host galaxy SED using data obtained from various archival observations. The photometric data encompassed $u$-, $g$-, $i$-, and $z$-band data (cmodel mags) from the Sloan Digital Sky Survey Data Release 17\footnote{https://www.sdss4.org/dr17/} \citep{abdurrouf22}, $g$-, $r$-, $i$-, $z$-, and $y$-band data (Kron mags) from the Pan-STARRS survey catalog\footnote{https://catalogs.mast.stsci.edu/panstarrs/} \citep{chambers16}, as well as J-, H-, and K-band data from the Two Micron All Sky Survey\footnote{https://www.ipac.caltech.edu/2mass/releases/allsky/} (2MASS), and the $W1$-, $W2$-, $W3$-, and $W4$-band data from AllWISE Source Catalogue\footnote{https://wise2.ipac.caltech.edu/docs/release/allwise/} \citep{wright10,mainzer11}.

The collected data is not simultaneous and originates from widely different observing epochs, potentially impacting the SED fitting process, particularly if the galaxy exhibits variability. The Pan-STARRS data comprises several stacked images (ranging from 9 to 23, depending on the band) with a mean epoch of 2013-01-20. In contrast, the SDSS data was acquired on 2001-05-23, significantly earlier. The 2MASS data predates 2001, obtained during the mission's operational lifetime, while the WISE data comprises three stacked epochs spanning 2010-2011.

To assess the variability of the host galaxy, we examined pre-transient archival survey photometry. Specifically, we obtained V-band data from the Catalina real-time transient survey (CRTS), covering the period from 06-15-2006 to 09-28-2013, W1 (3.4 micron) data from the NEOWISE mission spanning from 01-10-2010 to 06-15-2021, and G-band imaging from the Gaia mission spanning from 11-05-2014 to 05-23-2021. Visual inspection of this data revealed no significant large-scale flares resembling AT 2022fpx, nor did we observe any clear stochastic variations typically associated with AGN.

We applied a 3$\sigma$-clipping procedure to the measured CRTS magnitudes to identify any measurements that could indicate either variability or spurious readings. We found that the clipped measurements did not correspond to transient activity. Upon closer inspection, of the five values among the 166 measurements that were clipped, four exhibited significant decreases in magnitude, accompanied by larger uncertainties, suggesting inconsistency with genuine variability. The single measurement clipped due to an increase in magnitude was reinstated in the dataset. Subsequently, we binned the data to compare the measured magnitudes with the expected distribution. The standard deviation of the measured magnitudes was found to be consistent with the median photometric uncertainty of the individual measurements (Figure \ref{fig:CRTS}). Based on our analysis, we conclude that there is no evidence of V-band variability between 06-15-2006 and 09-28-2013. Consequently, it is improbable that the host optical magnitudes have varied between the SDSS and PanSTARRS epochs.

Performing a similar analysis for the NEOWISE and Gaia data poses more challenges due to the limited number of measurements available for our source (only 25 and 22 for NEOWISE and Gaia, respectively, compared to 162 for CRTS). Additionally, proper uncertainties for the Gaia measurements are unavailable as they are not provided by the Gaia Alerts service. Moreover, given that the host of AT 2022fpx is resolved at the Gaia resolution, variations in the measured magnitude could arise due to different scanning directions during observation.

In Fig. \ref{fig:NEOWISE}, we present the time evolution of the NEOWISE data, with the 1$\sigma$ region highlighted. While two data points exhibit potential connections to real variability at a 0.2 mag level, it's worth noting that our data originates from the AllWISE catalog, corresponding to the first three data points. Consequently, although we cannot definitively rule out variability in the host without examining the individual stacked images, the magnitudes we utilize align well with the mean galaxy magnitude.

\section{CIGALE fit parameters}

\begin{figure*} 
    \hbox{
    \includegraphics[width=0.5\textwidth]{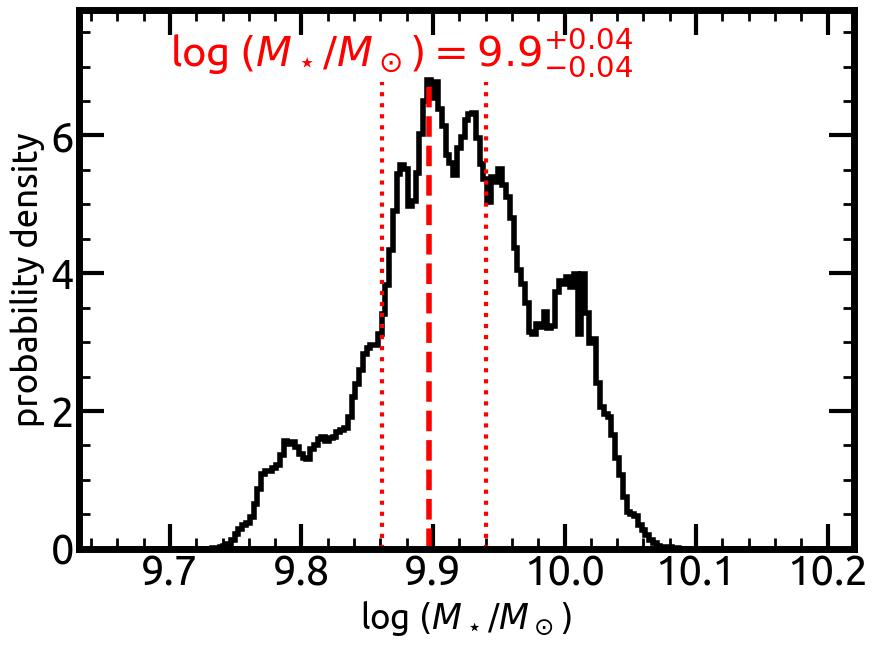}
    \includegraphics[width=0.5\textwidth]{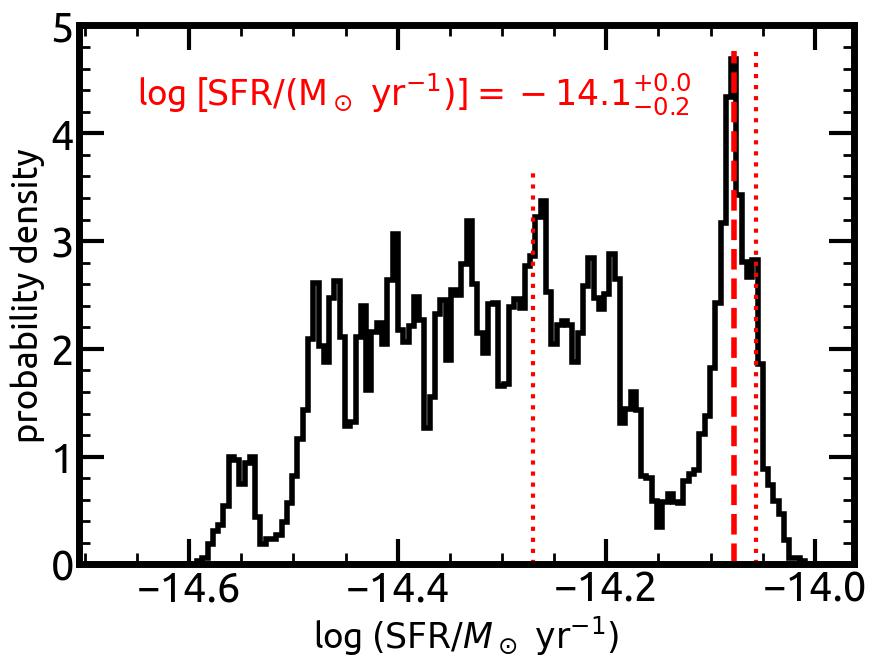}
    }
    \hbox{\includegraphics[width=0.5\textwidth]{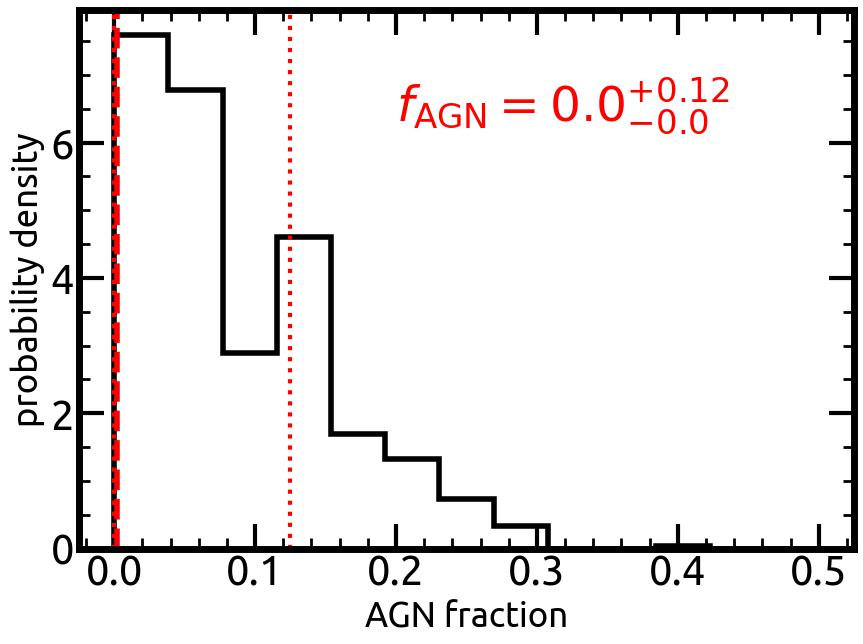}}
    \caption{The $M_\star$ (upper left), SFR (upper right), and AGN~fraction (bottom) $\chi^{2}_{\nu}$-weighted probability densities of all \texttt{CIGALE} models (black line). 
    The modes and 68\%/32\% confidence intervals are shown with vertical dashed and dotted red lines respectively.
    The estimated quantities are shown at the top of each panel with red fonts.}
    \label{fig:Mstar_SFR_PDFs}
\end{figure*}

The configuration of the \texttt{CIGALE} SED fitting modules is presented in Table \ref{tab:pcigale_ini}. 
A detailed description of the parameters is provided by \cite{boquien19}.
The $\chi^{2}_{\nu}$-weighted PDFs of the $M_\star$, SFR, and AGN~fraction, used to estimate the parameter's quantities and uncertainties, are shown in Fig. \ref{fig:Mstar_SFR_PDFs}.

\begin{table*}[ht]
    \caption{Configuration of the \texttt{CIGALE} SED fitting modules.}
    \begin{tabular}{l|c|r}
       \toprule
       Module & sfhdelayed   \\
       \hline
       tau\_main & 5, 10, 100, 200, 500, 1000, 3000, 5000 & e-folding time of the main SP model in Myr\\
       age\_main & 8000, 9000, 10000, 12000, 13000 & Age of the main SP in the galaxy in Myr \\
       tau\_burst & 1, 2, 5, 10, 25, 50, 100 & e-folding time of the late starburst population\\
       age\_burst & 100, 200, 500, 800, 900, 1000, 1100, 1500, 2500 & Age of the late burst in Myr\\
       f\_burst & 0.01, 0.1, 0.3, 0.4, 0.45, 0.5, 0.55, 0.6, 0.7, 0.8 & Mass fraction of the late burst population\\
       sfr\_A & 1 & Multiplicative factor controlling the SFR\\
       normalize & False & Normalize SFH to produce one solar mass\\
       \hline
       \hline
       Module & bc03\\
       \hline
       imf & 0 & Salpeter initial mass function\\
       metallicity & 0.0001, 0.0004, 0.004, 0.008, 0.02, 0.05 & Z\\
       separation\_age & 10 & Young and old SPs separation age in Myr\\ 
       \hline
       \hline
       Module & nebular \\
       \hline
       logU & -2.0, -1.0 & Ionization parameter\\
       f\_esc & 0.0 & Fraction of escaping $Ly$ continuum photons\\
       f\_dust & 0.0 & Fraction of absorbed $Ly$ continuum photons\\
       lines\_width & 300.0 & Line width (km/s)\\
       emission & True & Include nebular emission\\
       \hline
       \hline
       Module & dustatt\_modified\_starburst \\
       \hline
       E\_BV\_lines & 0.01, 0.1, 0.25, 0.3, 0.35, 0.4, 0.45, 0.5, 0.55, 0.6, 0.7 & color excess of the nebular lines\\
       E\_BV\_factor &  0.3, 0.35, 0.4 & Reduction factor applied on E\_BV\_lines.\\
       uv\_bump\_wavelength & $217.5$ & Central wavelength of the UV bump (nm)\\
       uv\_bump\_width & $35.0$ & Width (FWHM) of the UV bump (nm)\\
       uv\_bump\_amplitude & $0.0$ & Amplitude of the UV bump \\
       powerlaw\_slope & $-1.5$, $-0.5$, $0.0$, $0.5$ & Modifying slope $\delta$\\
       Ext\_law\_emission\_lines & 1 & 1 corresponds to MW extinction law\\
       $ R_{\rm V}$ & 3.1 & $A_{\rm V}/E(B-V)$\\
       \hline
       \hline
       Module & dale2014 \\
       \hline
       fracAGN & 0.0, 0.025, 0.05, 0.075, 0.1, 0.125, 0.15, 0.175, 0.2, 0.25, 0.3, 0.4, 0.5 & AGN fraction \\
       alpha &  1.5, 2.0, 2.5 & Alpha slope \\
       \hline
       \hline
       Module & restframe\_parameters\\
       \hline
       beta\_calz94 & True & UV slope as in Calzetti et al. (1994)\\
       D4000 & True & As in Balogh et al. (1999)\\
       IRX & True & based on GALEX FUV and dust luminosity\\
       \hline
       \hline
       Module & redshifting \\
       \hline
       redshift & 0.073 \\
    \hline
       Module &  \\
    \hline
    additional\_error & 0.1 & Relative error added in quadrature to the uncertainties\\
    & & of the fluxes and the extensive properties\\
    \bottomrule
    \end{tabular}
    \label{tab:pcigale_ini}
\end{table*}

\bsp	
\label{lastpage}
\end{document}